\documentclass{article}

\usepackage{PRIMEarxiv}
\usepackage{moreverb}
\usepackage{relsize}
\usepackage{nicefrac}
\usepackage{bm}
\usepackage{amsmath}
\usepackage{amssymb}
\usepackage{setspace}
\usepackage{lineno}
\usepackage{booktabs}
\usepackage{multirow}
\usepackage{makecell}
\usepackage{dsfont}
\usepackage{bbold}
\usepackage{soul}
\usepackage{color}
\usepackage{caption} 
\usepackage{mathrsfs}
\usepackage{optidef}
\usepackage{float}
\usepackage{comment}
\usepackage{hyperref}
\usepackage{algorithm} 
\usepackage{algpseudocode}
\usepackage{actuarialsymbol}
\usepackage{cleveref}

\newcommand*\samethanks[1][\value{footnote}]{\footnotemark[#1]}

\newcommand*{\affaddr}[1]{#1} 

\newcounter{algsubstate}


\usepackage{comment}
\usepackage{url} 
\usepackage{subcaption}
\usepackage{MnSymbol}%
\usepackage{wasysym}%

\title{GM-TOuNN: Graded Multiscale Topology Optimization using  Neural Networks}

\author{%
Aaditya Chandrasekhar \thanks{Contributed equally}, Saketh Sridhara \samethanks,   Krishnan Suresh \\
\affaddr{Department of Mechanical Engineering, University of Wisconsin-Madison}\\
\texttt{\{achandrasek3, ssridhara, ksuresh\}@wisc.edu%
}}

\begin{document}
\maketitle

\begin{abstract}
	Multiscale topology optimization (M-TO) entails generating an optimal global topology, and an optimal set of microstructures at a smaller scale, for a physics-constrained problem. With the advent of additive manufacturing, M-TO has gained significant prominence. However, generating  optimal microstructures  at various  locations can be computationally very expensive. As an alternate, \emph {graded multiscale topology optimization} (GM-TO) has been proposed where one or more pre-selected and graded (parameterized) microstructural topologies are used to fill the domain optimally. This leads to a significant reduction in computation while retaining many of the benefits of M-TO.

	A successful GM-TO framework must: (1) be capable of efficiently handling numerous pre-selected microstructures, (2) be able to continuously switch between these  microstructures during optimization, (3) ensure that the partition of unity is satisfied, and (4) discourage microstructure mixing at termination.
	
	In this paper, we propose to meet these requirements by exploiting the unique classification capacity of neural networks. Specifically, we propose a \emph {graded multiscale topology optimization using neural-network} (GM-TOuNN) framework with the following features: (1) the number of design variables is only weakly dependent on the number of pre-selected microstructures, (2) it guarantees partition of unity while discouraging microstructure mixing, and (3) it supports automatic differentiation, thereby  eliminating manual sensitivity analysis. The proposed framework is illustrated through several examples.

{\centering
\includegraphics[scale=0.6]{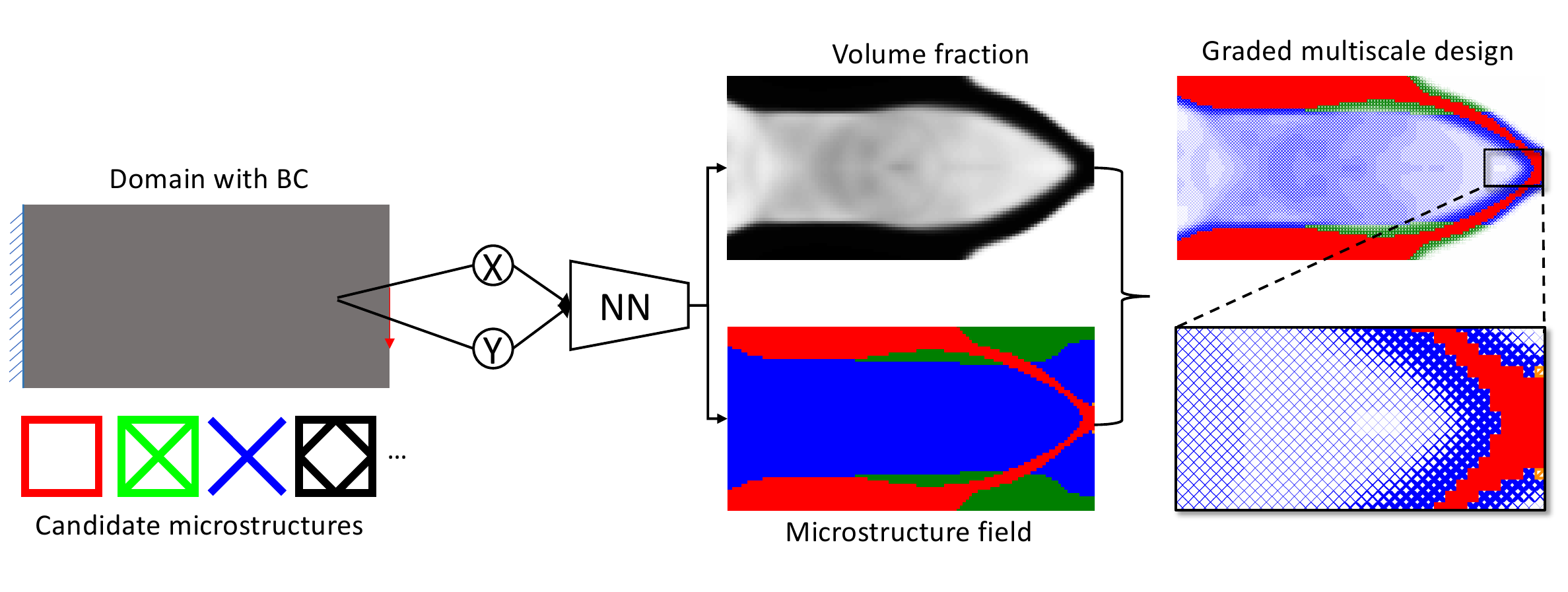}
\captionof*{figure}{Overview of the proposed method: Given candidate microstructures and a topology optimization problem, a neural network (NN) maps the spatial coordinates to a volume fraction field and a microstructure field, that produce a graded multiscale design.}
\par
}
\end{abstract}


\section{Introduction}\label{sec:intro}

\begin{figure*}[htbp]
	\begin{center}
		\includegraphics[scale=1.5,trim={0 0 0 0},clip]{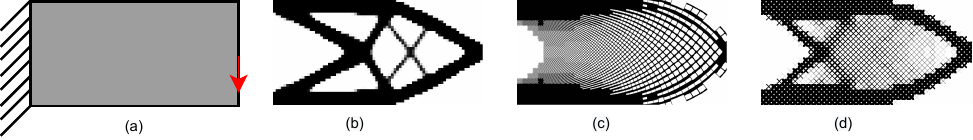}%
		\caption{TO problem and solutions: (a) Mid-cantilever problem. (b) Single-scale solution. (c) Multiscale solution. (d) Graded multiscale solution using a single X-shaped microstructure.}
		\label{fig:comparison_single_multiscale}
	\end{center}
\end{figure*}

Topology optimization (TO) is a strategy for  optimally distributing material within a design domain (\cite{Bendsoe2004,Sigmund2013}) to maximize a desired objective, while meeting one or more constraints. Various TO methods have been proposed; these include density-based methods(\cite{Bendsoe1989}), homogenization (\cite{Bendsoe1988,Hassani1998}), level-set (\cite{Sethian2000,Wang2003}), evolutionary (\cite{Xie1993,Yang1999}), morphable components (\cite{Zhanga}), and topological sensitivity based methods (\cite{Suresh2010,Deng2015}). These methods typically lead to a single-scale design where all geometric features are of similar size. As an example, for the cantilever problem posed in \cref{fig:comparison_single_multiscale}(a),  a single-scale TO that minimizes the compliance, subject to a volume constraint, leads to the topology in \cref{fig:comparison_single_multiscale}(b).

In contrast, in  multiscale topology optimization (M-TO),  geometric features are computed at two or more length scales (\cite{Rodrigues2002},\cite{Wang2017a}). For the problem in \cref{fig:comparison_single_multiscale}(a), \cref{fig:comparison_single_multiscale}(c) illustrates a topology, based on a two-scale M-TO (\cite{kumar2020density}). Such designs exhibit unique characteristics such as large surface-to-volume ratio, excellent resilience, etc. This  has led to a wide range of applications (\cite{Tao2016,Wang2018}) including energy absorption (\cite{Cheng2019}), heat exchangers (\cite{Li2019}), and biomedical applications (\cite{Liu2018e}). Current M-TO methods include clustering (\cite{Sivapuram2016,kumar2020density, jia2020multiscale}), kriging (\cite{Zhang2019}), multi-material (\cite{Zhang2018}), conformal lattices (\cite{wu2021efficient}), etc. The advent of additive manufacturing has further spurned research in M-TO (\cite{Tao2016}).

\subsection{Graded Multiscale TO }
\label{sec:litReview}

However, one of the challenges in M-TO is the high computational cost since one must generate and evaluate  various microstructures (through homogenization) during each step of the optimization process (\cite{Rodrigues2002,Sivapuram2016,kumar2020density,gu2022improved}). To reduce this cost, researchers have proposed the use of graded variations of one or more pre-selected microstructural topologies (\cite{Wang2017a,Li2019,thillaithevan2021stress}). \cite{Wang2017a} expressed the mechanical properties as a  density function, with a B-spline-based interpolation. \cite{Li2019} designed graded cellular structures With triply periodic shapes. \cite{thillaithevan2021stress} parameterized the radii of each member of a lattice with linear and sinusoidal graded variations. These methods are referred to as \emph {graded multiscale TO}, or GM-TO, leading to topologies such as the one illustrated in \cref{fig:comparison_single_multiscale}(d), where, as an example, graded variations of a single X-shaped microstructure are used everywhere. This leads to a significant reduction in computational cost (\cite{Wang2009}) since homogenization is not needed within the optimization loop. However, restricting the design to a single microstructure can significantly reduce performance (\cite{Wu2020}). To further improve the performance of GM-TO, without substantially increasing the computational cost, one can use \emph {a finite number of graded microstructures}. 

\textbf{Density-based methods:} Theoretically, using a finite number of microstructures in GM-TO is analogous to using multiple materials in TO. Therefore, various authors (\cite{Liu2020c, xu2017two, wang2019concurrent, zhou2022hierarchical}) proposed the use of uniform multi-phase materials interpolation (UMMI) models for GM-TO. Given the elasticity matrix of each microstructure at a particular volume fraction, two different UMMI models have been proposed for interpolation~\cite{Stegmann2005,Gao2011}.
The first type (UMMI-1) proposed by \cite{xu2017two,zhou2022hierarchical}, simply adds the contributions from each microstructure, and does not penalize microstructure mixing. The UMMI of the second type (UMMI-2) penalizes microstructure mixing at the cost of non-linearity (\cite{Liu2020c}). To mitigate the effect of non-linearity, \cite{Sanders2018} suggested a gradual increase in the penalty in multi-materials, further explored as a multi-scale formulation by \cite{sanders2021optimal,senhora2022optimally}.  Despite penalization, microstructure mixing is not entirely eliminated in UMMI-2. Further, while multiple materials can often co-exist at a given point, this is unacceptable in GM-TO. Finally, a handful (typically less than five) materials can be sufficient in multi-material TO, while a large number of microstructures are required in GM-TO. 

\textbf{Level-set methods:} Level-set methods are also a popular choice for GM-TO. 
\cite{nguyen2021multiscale,zhao2022stress} considered a level-set based method where the microstructural shape is represented and parametrized by implicit functions thereby circumventing the need for homogenization within every loop; instead one can rely on curve-fitting.  \cite{yu2022multiscale} utilized the level set function  to evolve on topologies on both the micro and macroscale, with connectivity ensured by the higher-order continuity of the level-set function. 
\cite{zhou2019level} addressed the challenge of connectivity of microstructures through shape metamorphosis to build graded transition zones using a connectivity index. Finally, 
 GM-TO with graded variations of a single microstructure was implemented by \cite{liu2019novel}  based on the parametrized level-set functions using radial basis functions.

\textbf{Data-driven methods:}
Recently, data-driven techniques have been proposed to address GM-TO. For instance,  \cite{wang2021data,wang2022data} utilize a latent variable Gaussian process that embeds discrete microstructures in a continuous and differentiable latent space. While the latent space is trained on the discrete homogenized data, the procedure of snapping and re-optimization may lead to sub-optimal results. Several authors (\cite{zheng2021data, Watts2019, White2019}) replaced the expensive homogenization process with neural networks that map the design variables to the homogenized elasticity matrix. \cite{zhou2022hierarchical} used neural networks to interpolate the stiffness matrices from homogenized data. \cite{patel2022improving} used two neural networks trained on topology optimized data: one for determining the microstructural topology and the other to improve the connectivity of microstructures. One of the challenges in data-driven methods is the computational expense in data generation and training of the machine learning models. Another challenge with the data-driven methods is that physical bounds on the elasticity matrices (such as positive-definiteness) cannot be enforced explicitly, which can lead to spurious stiffness matrices and non-convergence.
\subsection{Paper Contributions }
\label{sec:paperContributions}
In this paper, we exploit the unique classification capability of neural networks to address some of the limitations of current GM-TO methods. Specifically, we propose a \emph {graded multiscale topology optimization using neural-networks} (GM-TOuNN) (\cref{sec:methodology}) framework.  GM-TOuNN extends the mesh-independent neural-network
(NN) based representation of the macro-scale topology proposed by \cite{ChandrasekharTOuNN2020} to GM-TO. The mesh-independent representation allows us to consider a large number of candidate microstructures without increasing the number of design variables. Further, the partition of unity is implicitly guaranteed by the NN construction.  Finally, one can leverage automatic differentiation (AD) of the NN computational framework to provide end-to-end differentiability and automated sensitivity analysis. Numerical experiments are presented in \cref{sec:experiments}. We conclude with limitations and future work in \cref{sec:conclusion}.

\section{Proposed Method}
\label{sec:methodology}
\subsection{Problem Specification}
\label{sec:method_problemSpec}

\begin{figure}[htbp]%
	\begin{center}
		\includegraphics[scale=2,trim={0 0 0 0},clip]{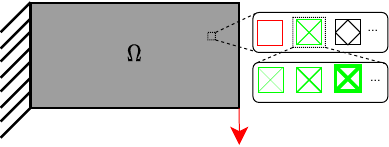}%
		\caption{A GM-TO problem.}
		\label{fig:proposedProblem}
	\end{center}
\end{figure}

Consider a design domain with prescribed loads, restraints and a set of microstructural topologies (\cref{fig:proposedProblem}). In the current work, we seek to compute an optimal multiscale design where we determine, at every location, the appropriate microstructure and its size-parameter (gradation). For simplicity, we will assume the compliance must be minimized subject to a volume constraint.
 
To assist in optimization, we introduce the following design variables. The presence or absence of a microstructure $m$ at any point $\bm{x}$ will be denoted by the variable $\rho_m(\bm{x})$, where, ideally, $\rho_m(\bm{x}) \in \{0,1\}$. However, for continuous gradient-based optimization, we will let $0 \le \rho_m \le 1$ and drive it towards $0/1$ through penalization.  Thus, at any point $\bm{x}$, one can define the vector $\bm{\rho}(\bm{x}) = \{\rho_1(\bm{x}), \rho_2(\bm{x}), \ldots, \rho_M(\bm{x})\}$  that captures the presence or absence of the $M$ microstructures with the \emph {partition of unity} constraint $\sum \rho_m(\bm{x}) = 1$. Further, since the microstructures are graded, we control their size by the scalar design variable $0 \le v(\bm{x}) \le 1$ where 0 denotes a void and 1 denotes a complete fill. Thus, in conclusion, $\bm{\rho}(\bm{x})$ dictates the type of microstructure, while $v(\bm{x})$ dictates the size or volume fraction of the microstructure at $\bm{x}$.

Consequently, one can pose the GM-TO problem in a discrete finite-element setting as:
\begin{subequations}
	\label{eq:optimization_base_Eqn}
	\begin{align}
		& \underset{\bm{\rho}, \bm{v}}{\text{minimize}}
		& &J(\bm{\rho}, \bm{v}) = \bm{f}^T \bm{u} \label{eq:optimization_base_objective}\\
		& \text{subject to}
		& & \bm{K}(\bm{\rho}, \bm{v})\bm{u} = \bm{f}\label{eq:optimization_base_govnEq}\\
		& & & g_v (\bm{v})  \coloneqq \frac{\sum\limits_e v(\bm{x}_e) A_e}{V_f^*\sum\limits_e A_e} - 1 \leq 0  \label{eq:optimization_base_volCons} \\
		& & & \sum\limits_{m=1}^M \rho_m(\bm{x}) = 1 \; , \; \forall \bm{x} \label{eq:optimization_base_partitionUnity}\\
		& & & 0 \leq \rho_m(\bm{x}) \leq 1 \; , \; \forall \bm{x} \; , \; \forall m \label{optimization_base_boundConsRho} \\
		& & &  0 \leq  v(\bm{x}) \leq 1 \; , \; \forall \bm{x} \label{optimization_base_boundConsV} 
	\end{align}
\end{subequations}
where $J$ is the structural compliance, $\bm{K}$ is the finite element stiffness matrix, $\bm{u}$ is the displacement field, $\bm{f}$ is the applied force, $V_f^*$ the maximum allowed volume fraction, with $A_e$ is the area (2D) of the element $e$.

\subsection{Design Representation using Neural Networks}
\label{sec:method_DesignRepresentationNN}

Typically, for such problems, the design variables are captured via the underlying FE mesh (\cite{Sanders2018Multimaterial}), i.e., for the above problem, design variables $\bm{\rho}_e$  and $v_e$ are defined at each element $e$. Thus the number of design variables grows  linearly with the mesh size. Further, observe that the partition of unity constraint must be imposed over each element.

\begin{figure}[htbp]%
	\begin{center}
		\includegraphics[scale=1,trim={0 0 0 0},clip]{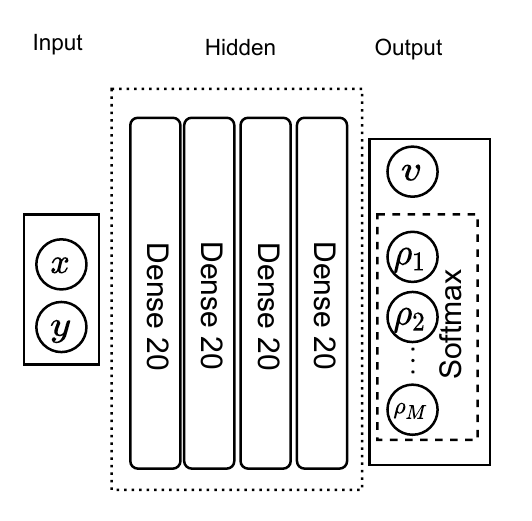}%
		\caption{Neural network architecture used for GM-TOuNN.}
		\label{fig:nnArchitecture}
	\end{center}
\end{figure}

In this paper, we avoid this undesirable growth in complexity by indirectly controlling the design variables via a coordinate-based neural-network (\cite{ChandrasekharTOuNN2020, chandrasekhar2021multi}). The proposed neural-network (NN) architecture (see \cref{fig:nnArchitecture}) consists of the following entities:

\begin{enumerate}
	\item \textbf{Input Layer}: The input to the NN are points  $\bm{x} \in \mathbf{R}^d$ ($d=2$ for 2D) within the domain. 
	\item \textbf{Hidden Layers}: The hidden layers  consists of a series of Swish ($x.sigmoid(x)$, \cite{ramachandran2017SwishActivation}) activated dense layers. 
	\item \textbf{Output Layer}: The output layer consists of $M+1$ neurons corresponding to the microstructure composition $\bm{\rho} = \{\rho_1, \rho_2, \ldots \rho_M\}$ and the volume fraction $v$. Further, the neurons associated with  $\bm{\rho}$ are activated by a softmax function. This guarantees that the partition of unity ($\sum \rho_i = 1$) and physical validity $0 \le \rho_i \le 1$ are automatically satisfied.  The output neuron associated with volume fraction is activated via a Sigmoid function, ensuring that $0 \le v(\bm{x}) \le 1$. Thus, no additional constraint are needed.
	\item \textbf{Design Variables}:  The weights and bias, denoted by the $\bm{w}$, now become the primary design variables, i.e., we have $\bm{\rho}(x,y; \bm{w})$ and $v(x,y; \bm{w})$. 
\end{enumerate}

Thus the strategy is to perform GM-TO via the NN weights $\bm{w}$, i.e., the
GM-TO problem in \cref{eq:optimization_base_Eqn} reduces to:
\begin{subequations}
	\label{eq:optimization_nn_Eqn}
	\begin{align}
		& \underset{\bm{w}}{\text{minimize}}
		& &J(\bm{w}) = \bm{f}^T \bm{u}(\bm{w}) \label{eq:optimization_nn_objective}\\
		& \text{subject to}
		& & \bm{K}(\bm{w}) \bm{u} = \bm{f}\label{eq:optimization_nn_govnEq}\\
		& & & g_v (\bm{w})  \coloneqq \frac{\sum\limits_e v(\bm{x}_e; \bm{w}) A_e}{V_f^*\sum\limits_e A_e} - 1 \leq 0  \label{eq:optimization_nn_volCons}
	\end{align}
\end{subequations}
Observe that: \emph {(1) no additional constraint is needed since they are automatically satisfied by the NN, and (2) increasing the number of candidate microstructures only increases the number of output neurons but not the number of design variables.}

\subsection{Material Model}
\label{sec:method_materialModel}

\begin{figure}[htbp]%
	\begin{center}
		\includegraphics[scale=0.35,trim={0 0 0 0},clip]{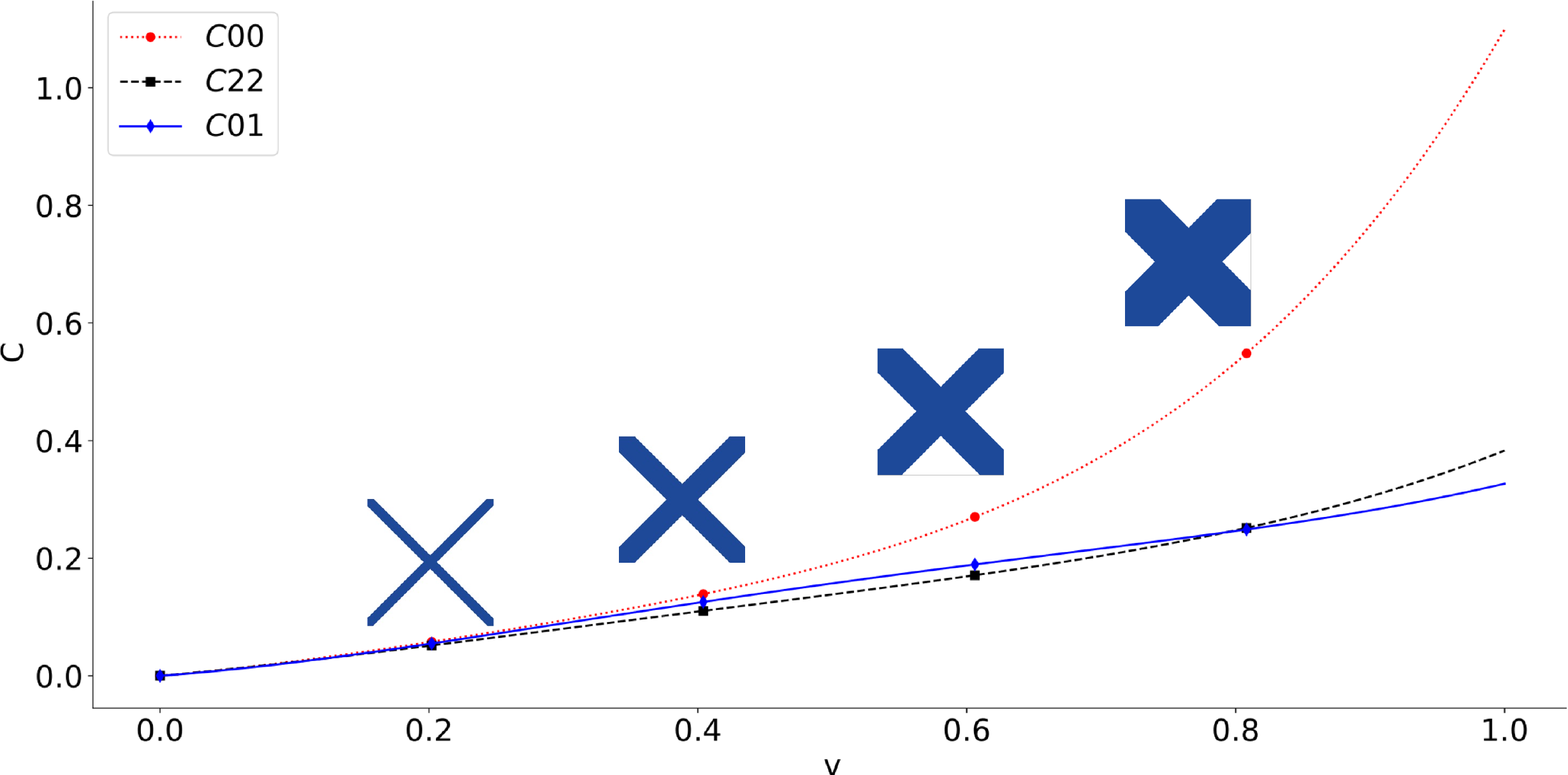}%
		\caption{Polynomial fit of the components of constitutive matrix for the $X$ type microstructure (note: $C_{00} = C_{11}$)} 
		\label{fig:XmstrCMatrixInterp}
	\end{center}
\end{figure}

Given the NN-architecture, one can now proceed to construct the material model for analysis. Towards this end, let $[C_m(v)]$ be the elasticity matrix of microstructure $m$ at volume fraction $v$, where $[C_m(v)]$ consists of six independent components (in 2D):
\begin{equation}
	[C_m(v)] = 
\begin{bmatrix}
  C_{11}(v) & C_{12}(v)  & C_{13}(v) \\
    & C_{22}(v)  & C_{23}(v) \\
    \multicolumn{2}{c}{\text{sym.}} &  C_{33}(v) 
\end{bmatrix}
\label{eq:CMatrixComponents}
\end{equation}
To obtain $[C_m(v)]$, we adopt a simple  constrained polynomial scheme  (\cite{sakethSpectral}). Specifically, the homogenized constitutive matrix \cite{andreassen2014Homogenization} for microstructure $m$ is evaluated at a few instances of volume fractions. Then, a polynomial ensuring positive definiteness is interpolated to these instances for each of the components. As an example, the polynomials for an X-type microstructure are illustrated in \cref{fig:XmstrCMatrixInterp}.

Given $[C_m(v)]$ for each microstructure $m$, the effective elasticity matrix  at any location $\bm{x}$ is defined here as a weighted average (\cite{Stegmann2005}):
\begin{equation}
    [C(\bm{\rho}, v)] = \sum\limits_{m=1}^M \rho_m^p [C_m(v)]
    \label{eq:netCMatrix}
\end{equation}
where $p$ is the solid isotropic material with penalization (SIMP) constant. The penalization discourages intermediate volume fractions, i.e., discourages microstructure mixing.

\subsection{Finite Element Analysis}
\label{sec:method_FEA}

We will use conventional finite element analysis (FEA) as part of the framework. Here, we use a structured quadrilateral mesh due to its simplicity.  The element stiffness matrix is defined as:

\begin{equation}
[K_e] = \int\limits_{\Omega_e} [\nabla N_e]^T[C(\bm{x}_e)][\nabla N_e] d \Omega_e
\label{eq:K_matrix_FE}
\end{equation}
where $[\nabla N_e]$ is the gradient of the shape matrix, and $[C(\bm{x}_e)]$ is the  elasticity tensor evaluated at the center of the element. 

In a straightforward, but na\"ive, implementation, one would evaluate $\bm{\rho}$ and $v$ at the center of each element via the NN. Then, given the pre-computed polynomials  $[C_m(v)]$, one would then find the effective $[C]$ for that element via  \cref{eq:netCMatrix}. Finally the element stiffness matrices will be computed via  \cref{eq:K_matrix_FE}. Thus the element stiffness matrices must be computed for each element, and for each step of the optimization process. This can be computationally very expensive.  To significantly reduce the computation, we  exploit the concept of \emph {template stiffness matrices} proposed by \cite{kumar2020density}.

Observe that one can express the element stiffness matrix as follows:

\begin{equation}
\begin{aligned}
[K_e] &=  C_{11} [\hat{K}^{1}] + C_{22}  [\hat{K}^{2}] +  C_{33}  [\hat{K}^{3}] \\&+ C_{12}  [\hat{K}^{4}] +  C_{13}  [\hat{K}^{5}] + C_{23}  [\hat{K}^{6}]
\label{eq:stiffnessElem_D_times_Ktemplate}
\end{aligned}
\end{equation}
where,
\begin{equation}
 [\hat{K}^i] = \int\limits_{\Omega_e} [\nabla N_e] ^T[\hat{C}^i][\nabla N_e]  d \Omega_e, \quad i = 1,2, \ldots 6
 	\label{eq:K_matrix_hat_1}
 \end{equation}
and

\begin{align}
	[\hat{C}^1] = \begin{bmatrix}
		1 & 0 & 0 \\
		0 & 0 & 0 \\
		0 & 0 & 0 \\
	\end{bmatrix} \;  [\hat{C}^2] = \begin{bmatrix}
		0 & 0 & 0 \\
		0 & 1 & 0 \\
		0 & 0 & 0 \\
	\end{bmatrix} \quad \; [\hat{C}^3] = \begin{bmatrix}
		0 & 0 & 0 \\
		0 & 0 & 0 \\
		0 & 0 & 1 \\
	\end{bmatrix}  \nonumber\\ 
	[\hat{C}^4] = \begin{bmatrix}
	0 & 1 & 0 \\
	1 & 0 & 0 \\
	0 & 0 & 0 \\
\end{bmatrix}  \; 	[\hat{C}^5] = \begin{bmatrix}
0 & 0 & 1 \\
0 & 0 & 0 \\
1 & 0 & 0 \\
\end{bmatrix} \quad  \; 	[\hat{C}^6] = \begin{bmatrix}
0 & 0 & 0 \\
0 & 0 & 1 \\
0 & 1 & 0 \\
\end{bmatrix}
\end{align}

Thus the matrices $ [\hat{K}^i]$ are computed at the start of the optimization. Then,  during optimization, the element stiffness matrices are computed efficiently via \cref{eq:stiffnessElem_D_times_Ktemplate}. In other words, one would evaluate $\bm{\rho}$ and $v$ at the center of each element via the NN. Then, given the pre-computed polynomials  $[C_m(v)]$, one would then find the effective $[C]$ for that element (as before). But, to compute the element stiffness matrices, we will rely on \cref{eq:stiffnessElem_D_times_Ktemplate} This will be followed by the assembly of the global stiffness matrix $\bm{K}$. 

\subsection{Loss Function}
\label{sec:method_lossFunction}

We now consider solving the NN-based optimization problem in \cref{eq:optimization_nn_Eqn}. Neural networks are designed to minimize a loss function using well-known optimization techniques such as Adam procedure (\cite{Kingma2015ADAM}). We therefore convert the constrained minimization problem into a loss function minimization by employing a log-barrier scheme as proposed in \cite{kervadec2019constrlogBarrierOptML}. Specifically, the loss function is defined as

\begin{equation}
    L(\bm{w}) = J + \psi(g_v)
    \label{eq:lossFunction}
\end{equation}
where, 
\begin{equation}
    \psi_t(g) = \begin{cases}
    -\frac{1}{t} \log(-g) \; ,\quad g \leq \frac{-1}{t^2}\\
    tg - \frac{1}{t} \log(\frac{1}{t^2}) + \frac{1}{t} \; , \quad \text{otherwise}
    \end{cases}
\end{equation}
where the parameter $t$ is updated during each iteration, making the enforcement of the constraint stricter as the optimization progresses. 
\subsection{Sensitivity Analysis}
\label{sec:method_Sensitivity}

\begin{figure*}%
	\begin{center}
		\includegraphics[scale=1,trim={0 0 0 0},clip]{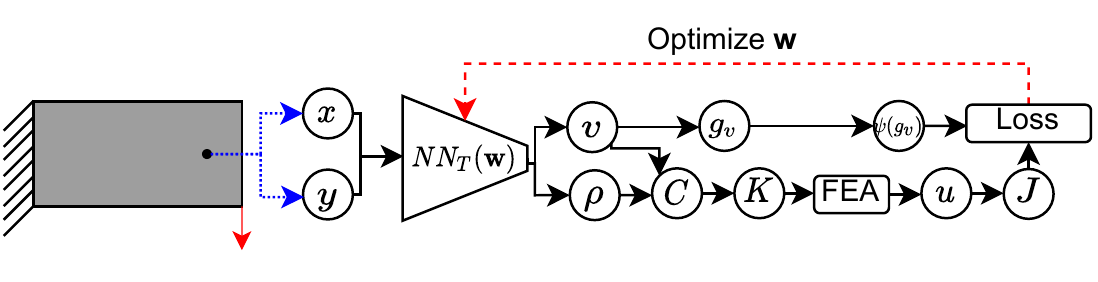}%
		\caption{Optimization loop of the proposed GM-TOuNN framework.}
		\label{fig:fwdComputation}
	\end{center}
\end{figure*}

Since Adam is a gradient-based optimizer, it requires sensitivities, i.e., derivative of the loss function (\cref{eq:lossFunction}) with respect to the design variable (weights of the NN $\bm{w}$). Fortunately, one can exploit modern automatic differentiation frameworks ( \cite{ chandrasekhar2021auto})  to avoid  manual sensitivity calculations. In particular, expressing all our computation, including FEA within JAX (\cite{jax2018github}), results in an end-to-end differentiable framework, as illustrated in \cref{fig:fwdComputation}. 

\subsection{Algorithm}
\label{sec:method_Algorithm}
We summarize the proposed framework in \cref{alg:GMTOuNN}. We will assume that a GM-TO problem, a desired volume fraction, an NN configuration and a set of microstructures with their interpolated coefficients are given. 

The first step in the algorithm is to discretize the domain with a finite element mesh, and sample the mesh (\cref{alg:elemCenterComp}) at the center of each element (these serve as inputs to the NN). We  compute the template stiffness matrices (\cref{alg:stiffnessTemplates}). The log barrier penalty $t$ and SIMP penalty parameter are also initialized (\cref{alg:Initialization}). 

In the main iteration, the element volume fraction $v$ and microstructure field $\bm{\rho}$ are computed using the NN using the current values of $\bm {w}$ (\cref{alg:fwdPropNN}). These fields are then used to construct the stiffness matrix and to solve for the displacement (\cref{alg:materialModel} - \cref{alg:feSolve}). Then we compute the  objective (\cref{alg:objectiveComp}) and volume constraint (\cref{alg:matrixVolCons}), leading to the loss function (\cref{alg:lossCompute}). The weights $\bm {w}$ are then updated using the Adam optimization scheme (\cref{alg:adamStep}). The optimizer requests the sensitivities which are computed in an automated fashion (\cref{alg:autoDiff}). Finally the log-barrier penalty and SIMP penalty parameters are updated (\cref{alg:OptPenaltyUpdate} and \cref{alg:simpPenaltyUpdate}). The process is  repeated until termination, i.e., till the relative change in loss is below a certain threshold or the iterations exceed a maximum value. Upon termination, each element is replaced with an image of the associated microstructure with the desired volume fraction. The framework is schematically depicted in \cref{fig:fwdComputation}.

\begin{algorithm}
	\caption{Graded Mutliscale TO}
	\label{alg:GMTOuNN}
	\begin{algorithmic}[1]
		\Procedure{GMTOuNN}{$\Omega_0$, BC, $V_f^*$, \ldots} \Comment{Inputs}
		
		\State $\Omega^0 \rightarrow \Omega^0_h$ \Comment{Discretize domain for FE} \label{alg:domainDiscretize}
		
		\State $\bm{x} = \{x_e,y_e\}_{e \in \Omega^0_h} $ \Comment{elem centers; NN input} \label{alg:elemCenterComp}

		\State $\Omega^0_h \rightarrow [\hat{K}]$ \Comment{Compute stiff.  templates \cref{eq:K_matrix_hat_1}} \label{alg:stiffnessTemplates}

		\State  k = 0; $t = t_0$; $p = 1$ \Comment{Initialization}
		\label{alg:Initialization}
		
		\Repeat \Comment{Optimization (Training)}
		
		\State $NN(\bm{x}_e ; \bm{w}) \rightarrow \{\bm{\rho}, v\}_e $ \Comment{Fwd prop NN} \label{alg:fwdPropNN}

        \State $v \rightarrow [C_m] \; m = 1, \ldots ,M \; \forall e$\Comment{elasticity matr. \cref{eq:CMatrixComponents}} \label{alg:materialModel}

        \State $ \{[C_m], \rho_m \}_e \rightarrow [C]_e $ \Comment{net elasticity matrix \cref{eq:netCMatrix}} \label{alg:netCMatrix}

		\State $\{[C], [\hat{K}]\}_{\forall e} \rightarrow [K]_e$ \Comment {elem stiff matrices \cref{eq:stiffnessElem_D_times_Ktemplate}} \label{alg:CMatrixToStiffness}

		\State $ \bigcup\limits_e [K]_e\rightarrow [K]$ \Comment {global stiff matrix} \label{alg:GlobalStiffness}
		
		\State $\{[K], \bm{f}\} \rightarrow \bm{u}$ \Comment{ FEA \cref{eq:optimization_nn_govnEq}} \label{alg:feSolve}
		
		\State$\{\bm{f}, \bm{u}\} \rightarrow J$ \Comment{Objective, \Cref{eq:optimization_nn_objective}} \label{alg:objectiveComp}

		\State $\{v_e, V_f^* \} \rightarrow g_v$ \Comment{ vol cons, \cref{eq:optimization_nn_volCons}} \label{alg:matrixVolCons}
		
		\State $\{J, g_v\} \rightarrow L$ \Comment{Loss from \Cref{eq:lossFunction}} \label{alg:lossCompute}
		
		\State $AD(L \leftarrow \bm{w}) \rightarrow \nabla L $ \Comment{sensitivity analysis} \label{alg:autoDiff}
			 
		\State $w  +  \Delta w (\nabla L ) \rightarrow w $ \Comment{Adam optimizer \cite{Kingma2015ADAM} step}\label{alg:adamStep}
		
		\State $\text{k}++$
		
		\State $ t  = t_0\mu^{k}$ \Comment {Increment $t$; log-barrier term} \label{alg:OptPenaltyUpdate}
		
		\State $p = p + \Delta p$ \Comment{SIMP penalty update}\label{alg:simpPenaltyUpdate}
		
		\Until{ $|| \Delta L || < \epsilon^*$ } and $k < k_{max}$ \Comment{convg. criteria}
		
	    \State	\Return $\bm{w}, \bm{\rho}, \bm{v}$

		\EndProcedure
	\end{algorithmic}
\end{algorithm}

\section{Numerical examples}
\label{sec:experiments}
\begin{figure*}[htbp]%
	\begin{center}
		\includegraphics[scale=0.5,trim={0 0 0 0},clip]{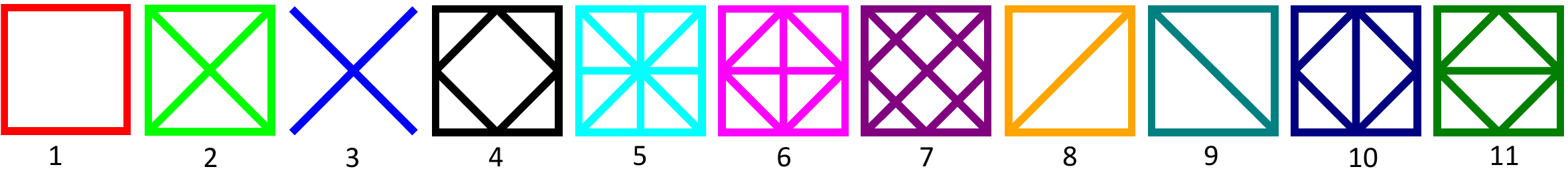}
		\caption{Microstructures considered in the experiments.}
		\label{fig:microstructuresUsed}
	\end{center}
\end{figure*}

In this section, we conduct several experiments to illustrate the method and algorithm. The default settings are as follows:

\begin{itemize}
    \item \textbf{Mesh}: A mesh size of $60 \times 30$ with elements of size $1 \times 1$ is used for all experiments, unless otherwise stated; the force is assumed to be 1 unit.
    
    \item \textbf{Material}: The microstructures are assumed to be composed of an isotropic material with a Poisson's ratio $\nu = 0.3$ and  Young’s modulus $E = 1$.
    
    \item \textbf{Neural Network}: The NN comprises of 4 Swish-activated hidden layers with 20 neurons in each layer. 
    
    \item \textbf{Candidate Microstructures}: A set of 11 predefined microstructures (\cref{fig:microstructuresUsed}) are used in the experiments. Observe that connectivity between these microstructures is guaranteed. A quintic polynomial is used to interpolate the components of the constitutive matrix.
    
    \item \textbf{Material Penalization}: The SIMP penalization constant $p$ in \cref{eq:netCMatrix} is incremented every iteration by $0.02$ using the continuation scheme \cite{Sigmund1998NumericalInstabTO}, starting from a value of $1$, capped at $8$.  
    
    \item \textbf{Loss Function}: The constraint penalty values of $t_0 = 1$ and $\mu = 1.01$ is used.
    
    \item \textbf{Optimizer}: Adam optimizer with a learning rate of $0.01$ is used. Further, the gradients are clipped at a maximum norm of 1 to improve stability of convergence (\cite{gradientClippingML}). A maximum of 300 iterations were allowed with $\epsilon^* = 0.01$
    
    \item \textbf{Computing Environment}: All experiments are conducted on a MacBook M1 Pro, using the JAX (\cite{jax2018github}) environment. 
    
\end{itemize}

\subsection{Validation}
\label{sec:expts_validation}

\begin{figure}[htbp]
  \begin{subfigure}[b]{\linewidth}
  \centering
    \includegraphics[width=0.65\linewidth]{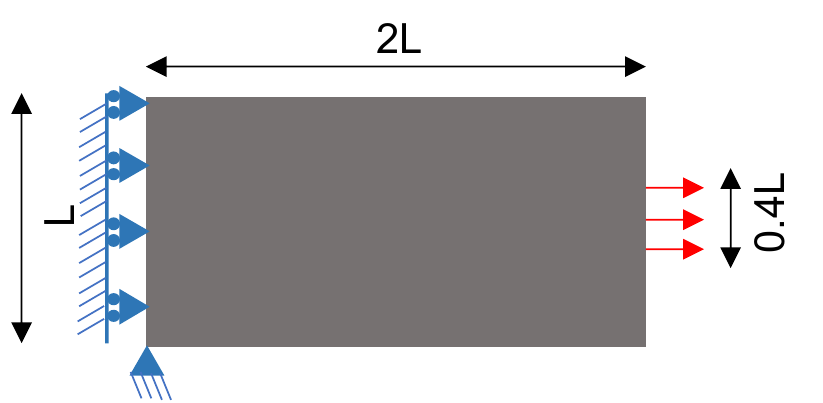}
    \caption{Loading of a tensile bar.}
    \label{fig:tensileBarBC}
  \end{subfigure}

  \begin{subfigure}[b]{\linewidth}
  \centering
    \includegraphics[width=0.4\linewidth]{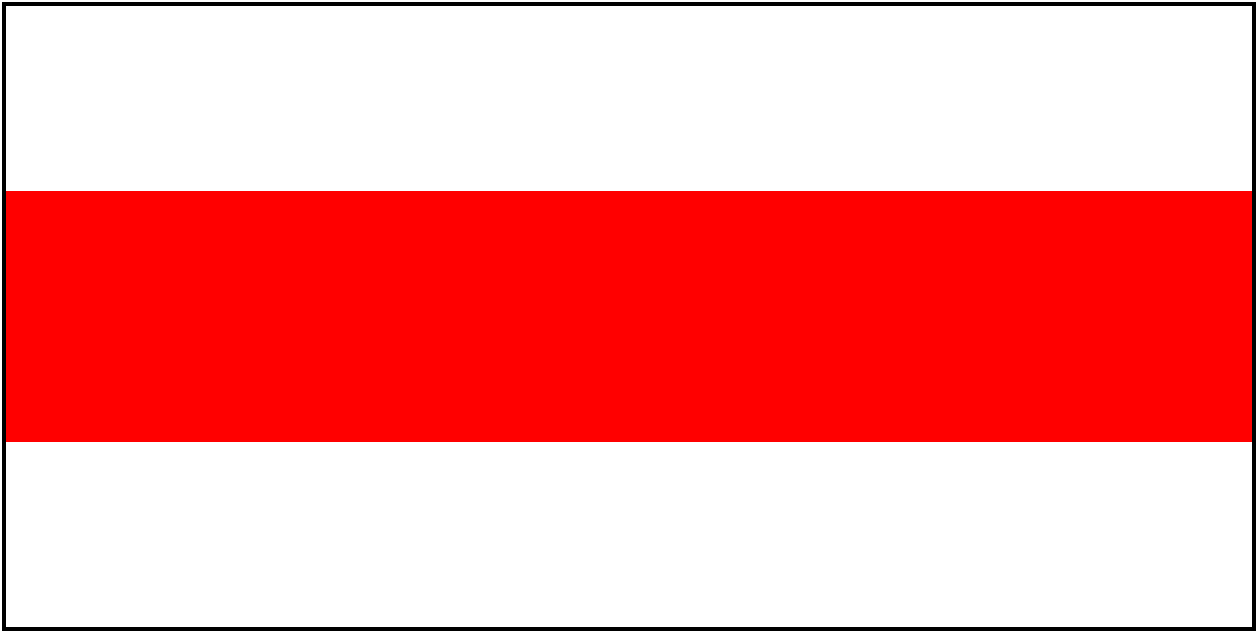}
    \caption{GM-TO of a tensile bar with $V_f^* = 0.4$ ($J^* = 183$).}
    \label{fig:tensileBar_40}
  \end{subfigure}
  \caption{A tensile-bar problem.}
  \label{fig:tensileBar}
\end{figure}

Consider the simple tensile bar problem in \cref{fig:tensileBarBC}. Although eleven microstructures were provided (see \cref{fig:microstructuresUsed}), the final topology in \cref{fig:tensileBar_40} consists of a single microtopology (number 1) with a complete fill for all elements in the middle and void at the top and bottom, i.e., the result is consistent with expectations.  The optimization converged in 240 iterations, taking 47 seconds.

\begin{figure}[htbp]
  \begin{subfigure}[b]{\linewidth}
  \centering
    \includegraphics[width=0.5\linewidth]{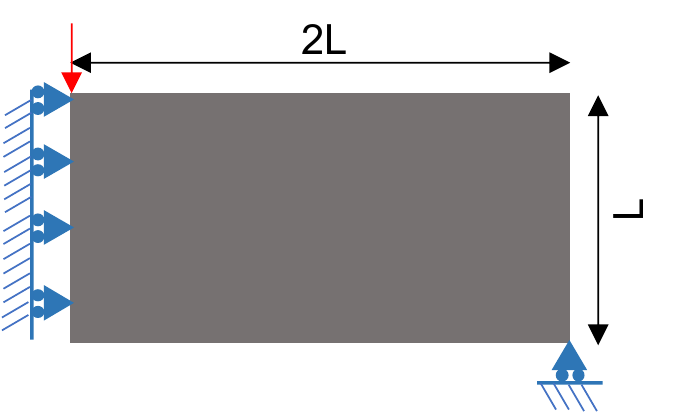}
    \caption{Loading of an MBB beam.}
    \label{fig:MBB_BC}
  \end{subfigure}

  \begin{subfigure}[b]{\linewidth}
  \centering
    \includegraphics[width=0.4\linewidth]{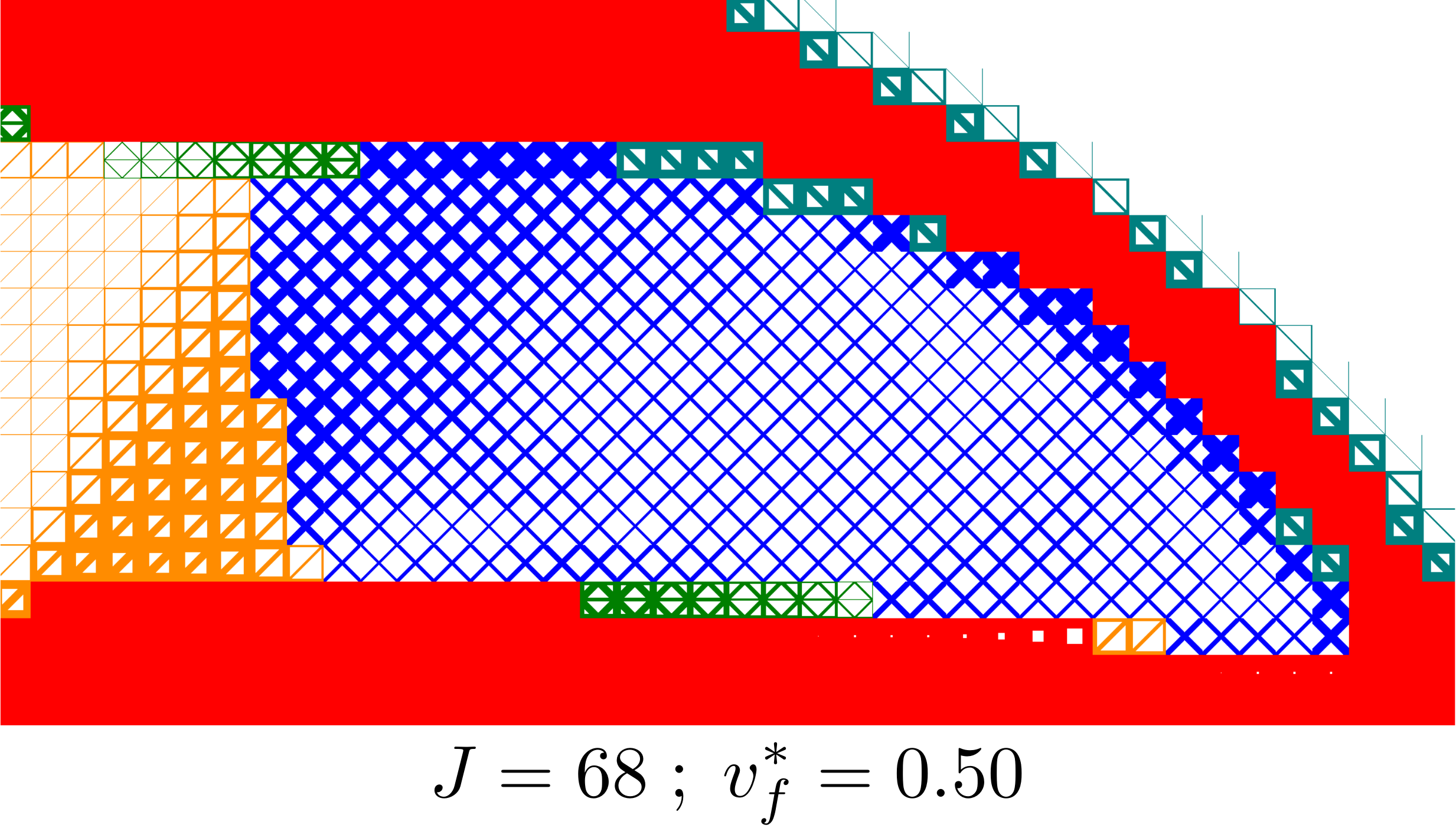}
    \caption{Optimized MBB beam.}
    \label{fig:MBB_50}
  \end{subfigure}
  \caption{GM-TO of an MBB beam.}
  \label{fig:MBB_validation}
\end{figure}

Next, we consider an MBB beam in \cref{fig:MBB_BC}.  \cite{sakethSpectral} reported compliance values between $76$ and $88$ when considering only one microstructure at a time. In comparison, we achieve a compliance of $68$ when all microstructures are allowed; see \cref{fig:MBB_50}. 

\begin{figure}[htbp]
	\begin{center}
		\includegraphics[scale=1,trim={0 0 0 0},clip]{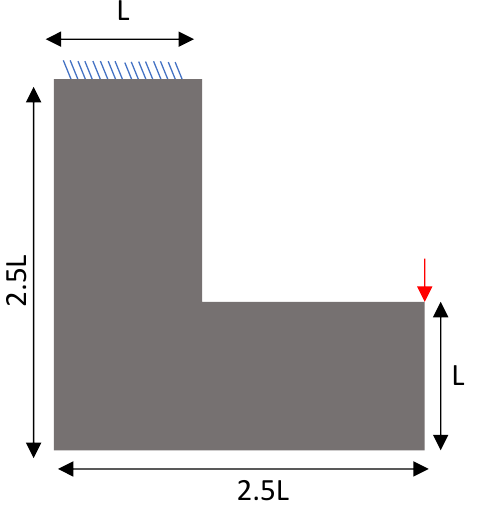}
		\caption{Loading of an L-bracket}
		\label{fig:LbracketBC}
	\end{center}
\end{figure}

\begin{figure*}[htbp]%
	\begin{center}
		\includegraphics[scale=0.45]{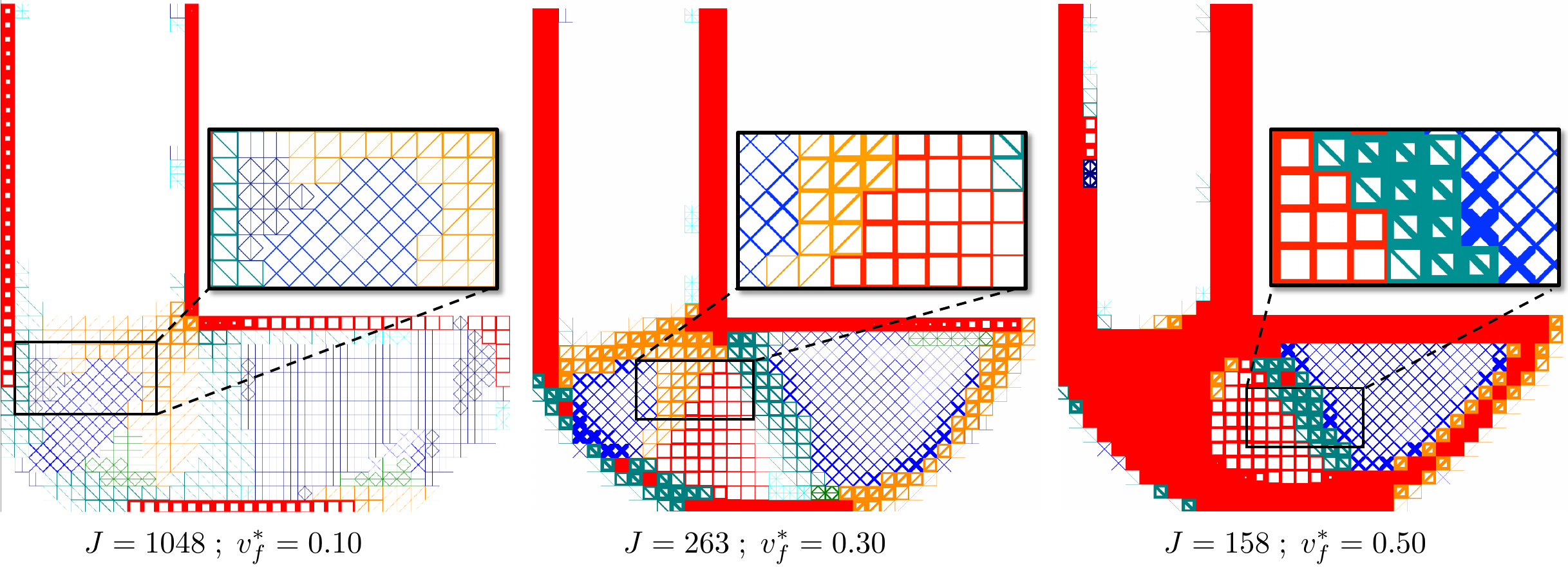}
		\caption{GM-TO of an L-bracket at various volume fractions.}
		\label{fig:LbracketPareto}
	\end{center}
\end{figure*}

Next, we consider an L-bracket (800 elements) illustrated in \cref{fig:LbracketBC}. Using only a single X-type microstructure, \cite{sakethSpectral} achieved a compliance of 1130, 283, 171 for volume fractions of 0.1, 0.3 and 0.5 respectively. By considering all microstructures (\cref{fig:LbracketPareto}), we achieved compliance of 1048, 263 and 158 respectively.

\begin{figure}[htbp]
  \begin{subfigure}[b]{\linewidth}
  \centering
    \includegraphics[width=0.45\linewidth]{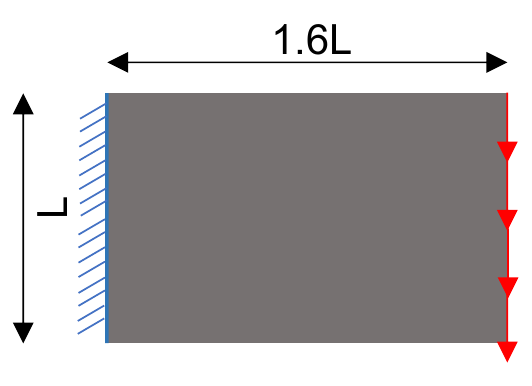}
    \caption{Edge-loaded beam.}
    \label{fig:fullLoadBC}
  \end{subfigure}

  \begin{subfigure}[b]{\linewidth}
  \centering
    \includegraphics[width=0.35\linewidth]{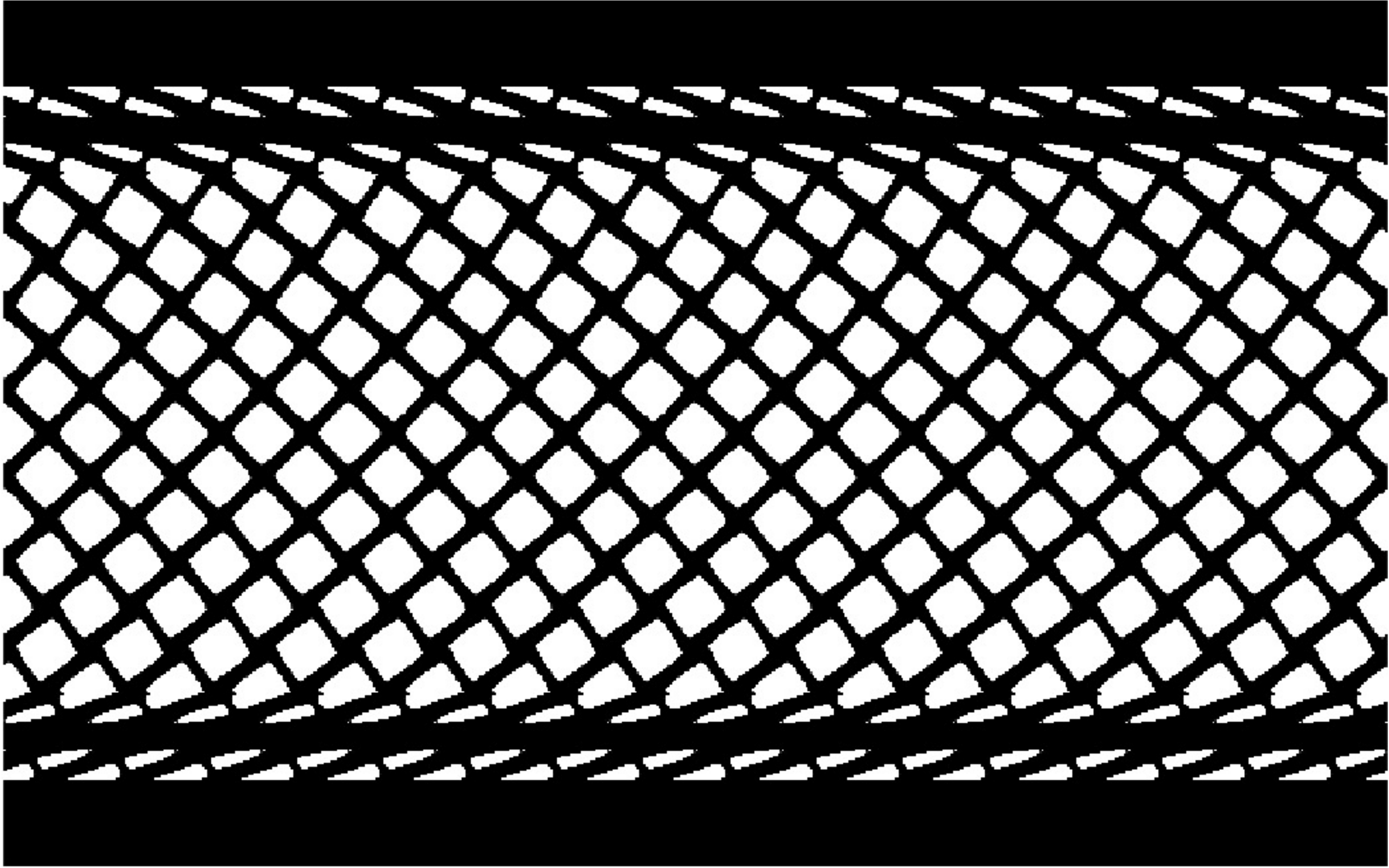}
    \caption{M-TO design ($J = 1.45e5$) obtained by \cite{kumar2020density}.}
    \label{fig:fullLoadTej}
  \end{subfigure}
  
\begin{subfigure}[b]{\linewidth}
  \centering
    \includegraphics[width=0.35\linewidth]{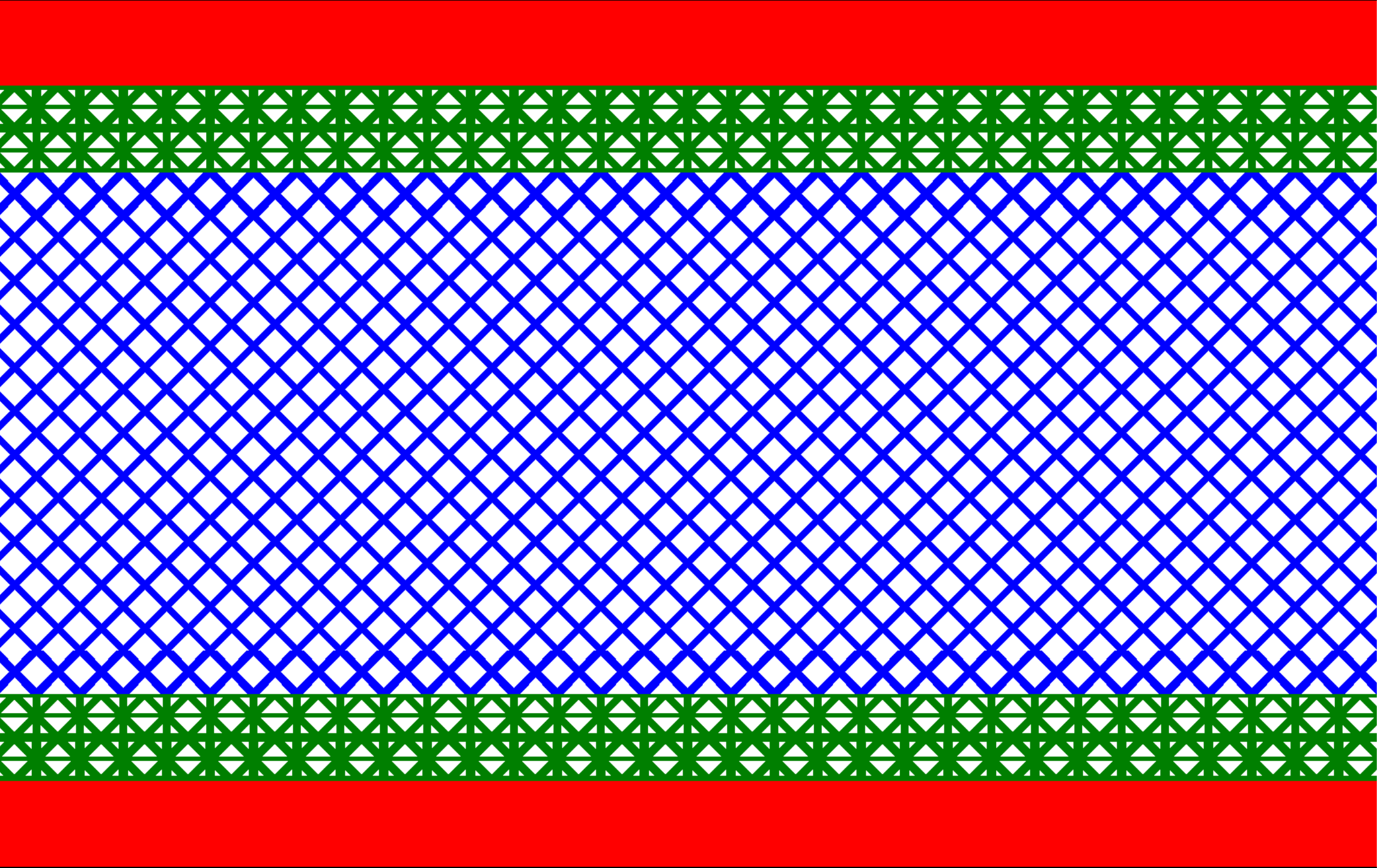}
    \caption{GM-TO design ($J = 1.41e5$)  obtained using the proposed method.}
    \label{fig:fullLoad}
  \end{subfigure}
  \caption{Validation for an edge-loaded beam.}
  \label{fig:edgeLoadValidation}
\end{figure}

Finally, we consider an edge-loaded beam with a mesh of $32 \times 20$ elements illustrated in \cref{fig:fullLoadBC}. Here, the design is constrained to have the same microstructure along the X-axis (but can vary along the Y-Axis), with a desired volume fraction $V_f^* = 0.6$. The topology and compliance obtained (\cref{fig:fullLoad}) is validated against the one reported by \cite{kumar2020density} (\cref{fig:fullLoadTej}). 

\subsection{Convergence Study}
\label{sec:expts_convergence}

\begin{figure*}[htbp]%
	\begin{center}
		\includegraphics[scale=0.5,trim={0 0 0 0},clip]{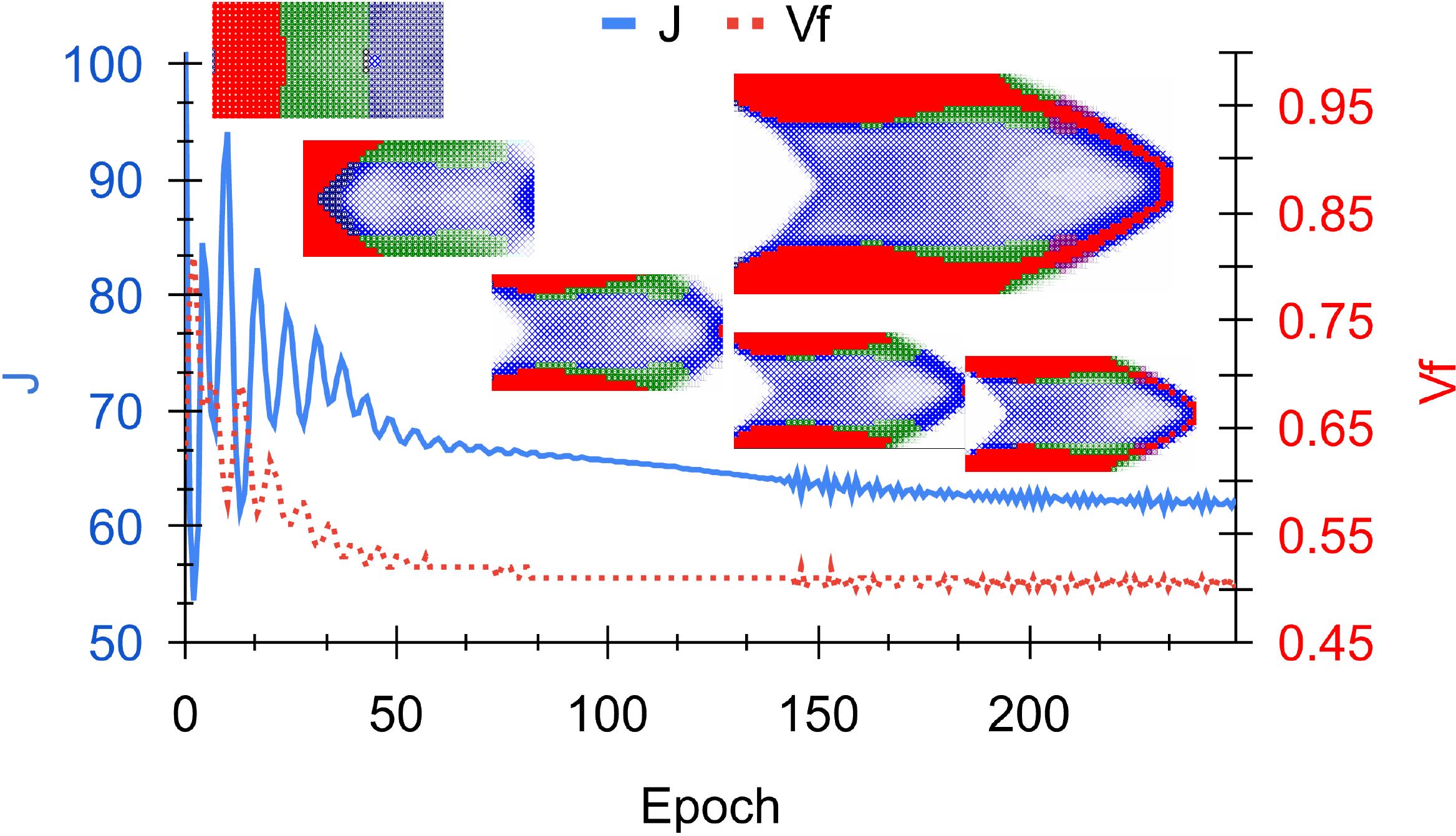}
		\caption{Convergence of compliance and volume fraction for a mid-cantilever beam. Topologies are illustrated at the $0^{\text{th}}$, $50^{\text{th}}$, $100^{\text{th}}$, $150^{\text{th}}$, $200^{\text{th}}$ and $250^{\text{th}}$(final) iterations.}
		\label{fig:convergence_midCant}
	\end{center}
\end{figure*}

We illustrate the typical convergence of the
proposed algorithm for a mid-cantilever beam (\cref{fig:comparison_single_multiscale}(a)) for $V_f^* = 0.5$. The compliance, volume fraction and the evolving topology at various instances are illustrated in \cref{fig:convergence_midCant}. We observe that the log-barrier formulation leads to a stable convergence. Similar convergence behavior was observed for all other examples. The computation took $53$ secs.

\subsection{Varying number of microstructures}
\label{sec:expts_numMicrostructures}

\begin{figure*}[htbp]
	\begin{center}
		\includegraphics[scale=0.5,trim={0 0 0 0},clip]{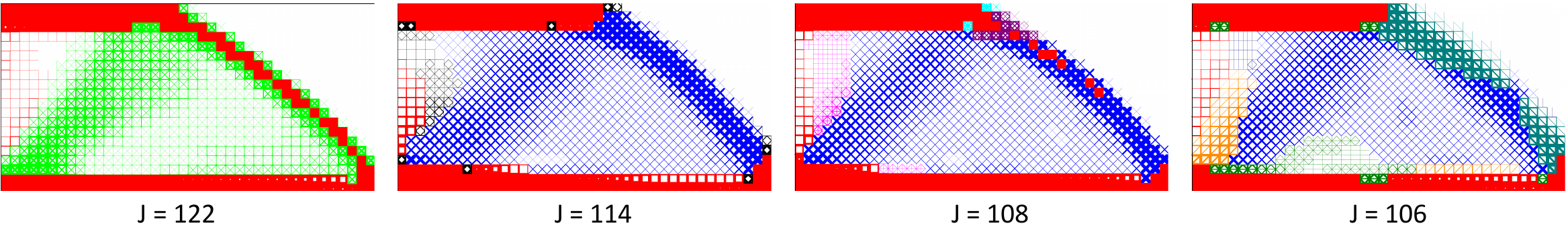}
		\caption{Design and compliance using 2, 4, 7 and 11 microstructures.}
		\label{fig:MBB_varyingMS}
	\end{center}
\end{figure*}

Two central hypotheses of the current work is that one can achieve better designs with larger number of candidate microstructures, and the framework is (computationally) insensitive to the number of candidates. To validate these, consider the problem in \cref{fig:MBB_BC}. The topologies and compliance values  when allowing for varying number of microstructures, in the order in which they are present in \cref{fig:microstructuresUsed}, are  illustrated in \cref{fig:MBB_varyingMS},  As expected, the compliance improves as we allow for larger number of microstructures. As noted earlier, each additional candidate microstructure requires only one additional output neuron (2 additional design variables), enabling us to consider large set of  microstructures if necessary. The computational time was approximately $56$ seconds, independent of the number of microstructures.

\subsection{Pareto Designs}
\label{sec:expts_pareto}

\begin{figure*}[htbp]
	\begin{center}
		\includegraphics[scale=0.4]{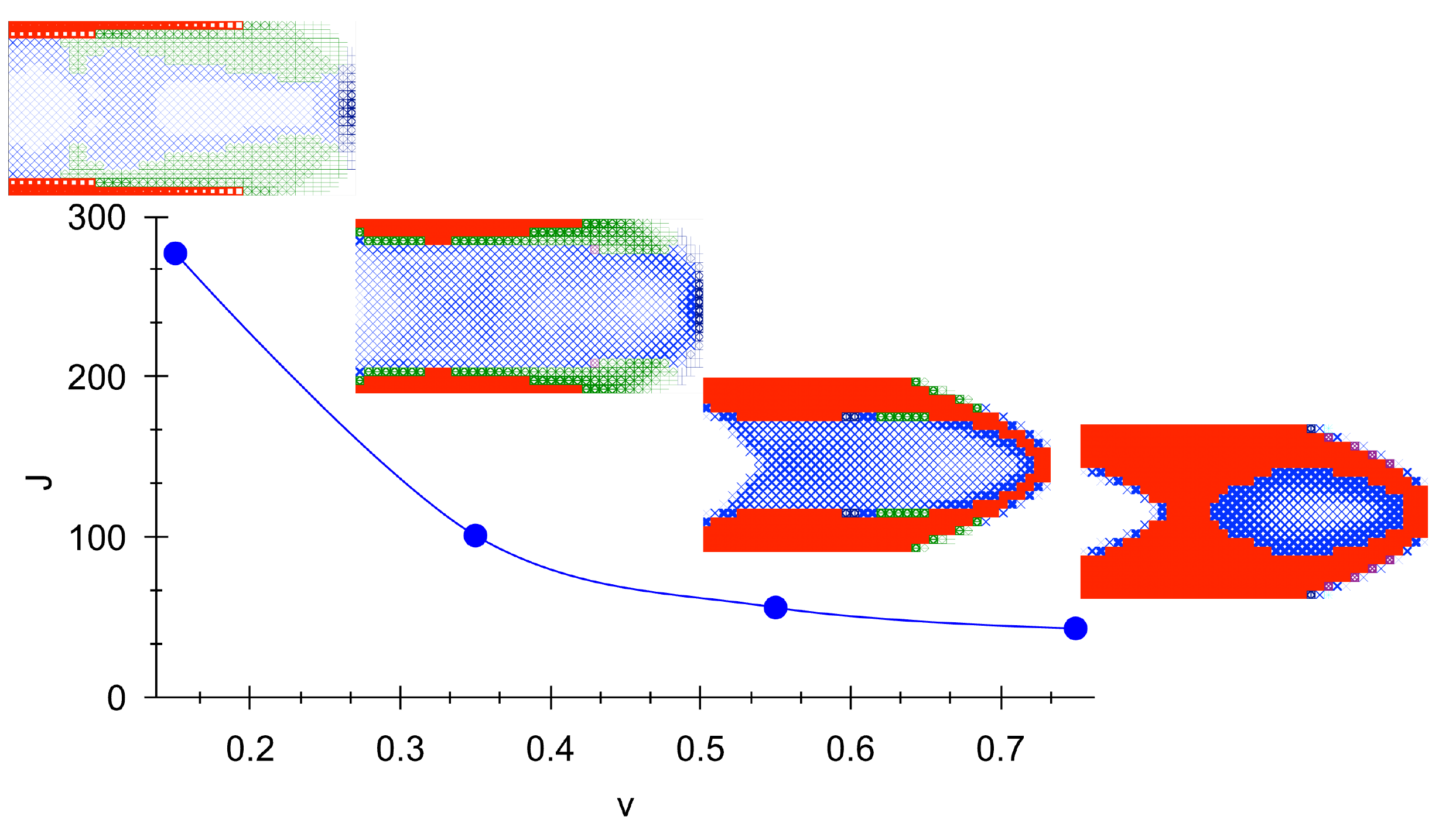}
		\caption{Pareto designs of mid-cantilever beam.}
		\label{fig:pareto_midCant}
	\end{center}
\end{figure*}

Exploring the Pareto-front is critical in making design choices and understanding the trade-off between the objective and constraint. Consider once again the mid-cantilever beam in \cref{fig:comparison_single_multiscale}(a). We illustrate the compliance values and topologies for varying volume fractions $V_f^*$ in \cref{fig:pareto_midCant}. 

\subsection{Mesh and NN Dependency}
\label{sec:NNSize}

\begin{figure*}[!htbp]
	\begin{center}
		\includegraphics[scale=0.3]{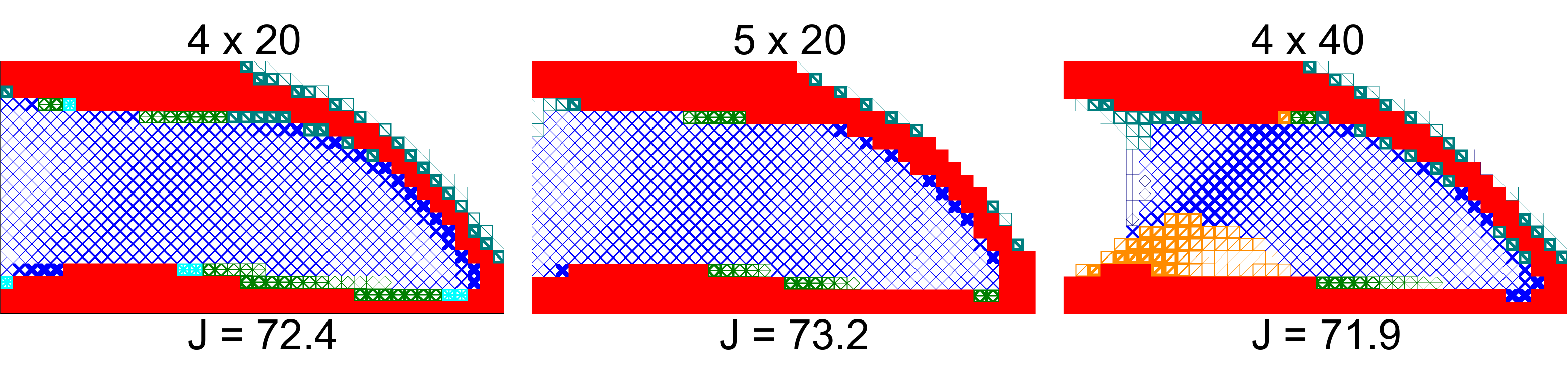}
		\caption{GM-TO of a tip cantilever at $V_f^* = 0.45$ with varying NN size.}
		\label{fig:NNDependency}
	\end{center}
\end{figure*}

\begin{figure*}[!htbp]
	\begin{center}
		\includegraphics[scale=0.3]{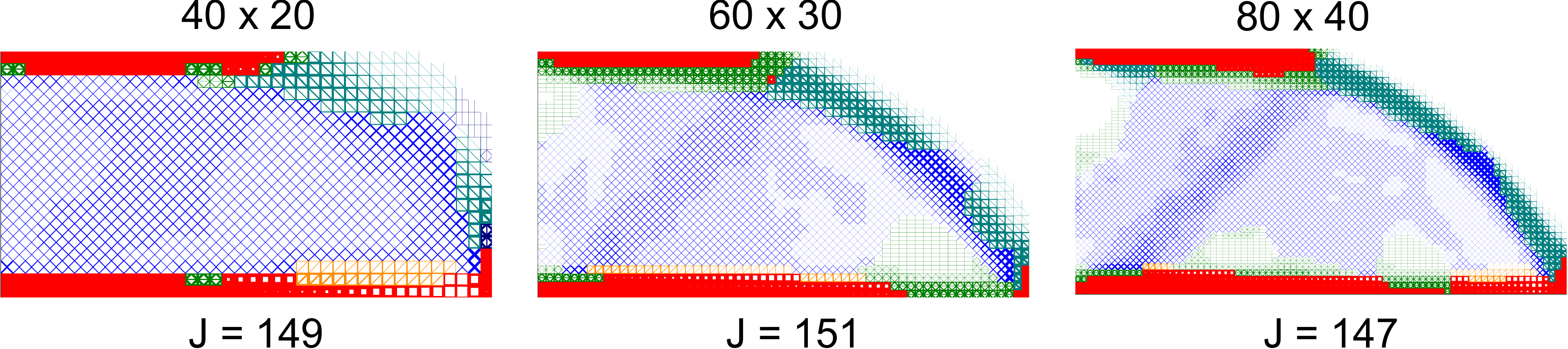}
		\caption{GM-TO of a tip cantilever at $V_f^* = 0.25$ with varying mesh size.}
		\label{fig:MeshDependency}
	\end{center}
\end{figure*}

Next we study the effect of NN size and FE mesh on the computed topology using the tip cantilever (\cref{fig:proposedProblem}) as an example; all other parameters are kept constant. For varying NN size, the topologies are reported in \cref{fig:NNDependency}; no appreciable difference in computational time was observed. Similarly, \cref{fig:MeshDependency} captures the topologies for varying mesh size; the computational times were 7.5, 24.7 and 114.3 seconds per 100 iterations for the $40 \times 20$, $60 \times 30$ and $80 \times 40$ mesh respectively.

\subsection{High Resolution Design}
\label{sec:expts_highResDesign}

\begin{figure*}[htbp]
	\begin{center}
		\includegraphics[scale=0.6,trim={0 0 0 0},clip]{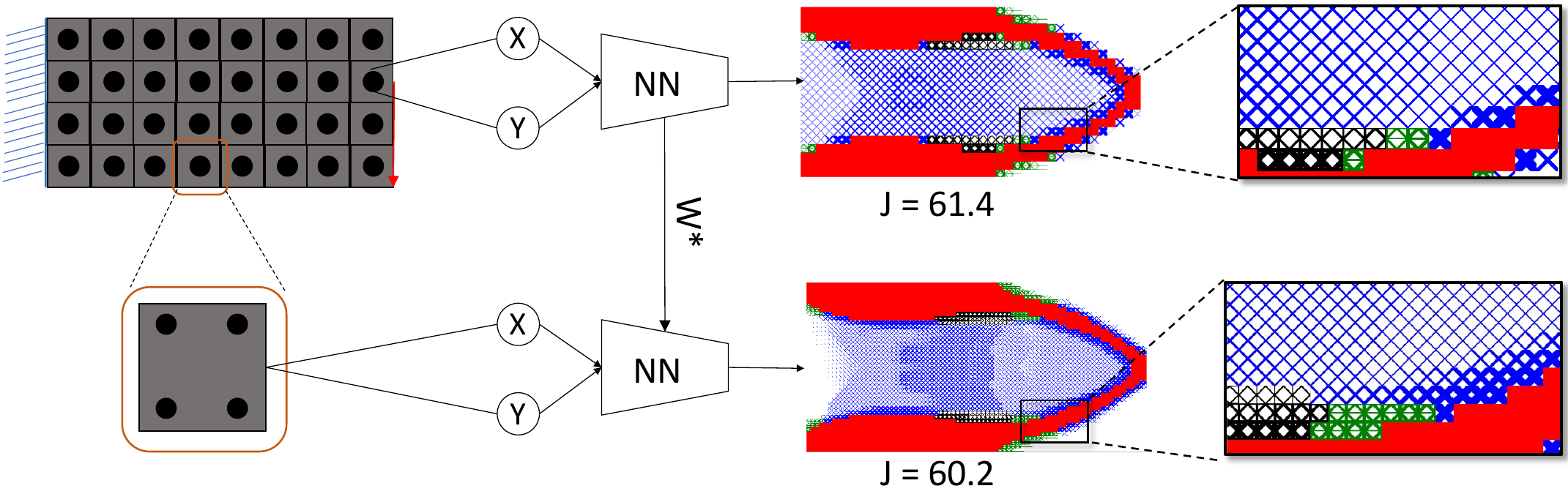}
		\caption{Extraction of high resolution design with smoother variation of microstructures.}
		\label{fig:highResDesign}
	\end{center}
\end{figure*}

In the proposed method, one can perform optimization on a coarse mesh and then re-sample at a higher resolution to populate the microstructures, resulting in smoother variation in the microstructure topologies.  Note that, even after re-sampling, the framework  guarantees partition of unity. However, due to NN-interpolation, microstructure mixing can occur on the finer mesh at the interfaces. This was resolved by choosing the microstructure $m$ with the largest $\rho_m$.  This is illustrated in \cref{fig:highResDesign}  where we optimize on a $40 \times 20 $ mesh and then re-sample on a $80 \times 40 $ mesh.

\section{Conclusions}
\label{sec:conclusion}

A framework for graded multi-scale topology optimization using neural network was proposed and demonstrated. The salient features of this framework are: (1) the number of design variables are only weakly dependent on the number of pre-selected microstructures, (2) the partition of unity constraint is automatically satisfied, and (3) manual sensitivity calculations is avoided. 

The framework was limited to 2D compliance minimization problems, involving microstructures governed by a single size parameter. Extension to 3D, non-compliance problems (such as energy absorption \cite{zhang2022uncover}, orthopedic implants \cite{ferro2022design}, resonant frequencies \cite{nightingale2021multiscale})  using more generic multi-parameter microstructures need to be explored. Furthermore, while we relied on polynomials for directly interpolating the elasticity components, it is desirable to consider an Eigen-value decomposition  (\cite{sakethSpectral}) for increased robustness.   The framework can complement and  might benefit from  data driven approaches (\cite{wang2021data, wang2022data}). Finally, while we relied on the simple Adam optimization procedure, second order optimization methods such as LBFGS might result in better/faster convergence  (\cite{nocedal2006numericalOptimization}).


\section*{Acknowledgements}
\label{sec:acknowledgements}

The authors would like to thank the support of  National Science Foundation through grant CMMI 1561899, and  the U. S. Office of Naval Research under PANTHER award number N00014-21-1-2916  through Dr. Timothy Bentley.

\section*{Replication of Results}
The Python code pertinent to this paper is available at \href{https://github.com/UW-ERSL/GMTOuNN}{github.com/UW-ERSL/GMTOuNN}

\section*{Compliance with ethical standards}
\label{sec:replicationResults}
The authors declare that they have no conflict of interest.


\bibliographystyle{unsrt}  
\bibliography{main_gmto}

\end{document}